\documentclass[journal=nalefd]{achemso}

\usepackage{graphicx} 
\usepackage{amsmath} 
\usepackage{amssymb} 

\usepackage{color}

\title{Strain induced bang-gap engineering in layered $\text{TiS}_3$.} 
\author{Robert Biele} 
\email{r.biele02@gmail.com}
\affiliation{Nano-Bio Spectroscopy Group and European Theoretical Spectroscopy Facility (ETSF), Universidad del Pa\'is Vasco, E-20018 San Sebasti\'an, Spain}

\author{Eduardo Flores}
\affiliation{Materials of Interest in Renewable Energies Group (MIRE Group), Dpto. de F\'isica de Materiales, Universidad Aut\'onoma de Madrid, UAM, 28049 Madrid, Spain}
\author{Jose Ram\'on Ares}
\affiliation{Materials of Interest in Renewable Energies Group (MIRE Group), Dpto. de F\'isica de Materiales, Universidad Aut\'onoma de Madrid, UAM, 28049 Madrid, Spain}
\author{Carlos Sanchez}
\affiliation{Materials of Interest in Renewable Energies Group (MIRE Group), Dpto. de F\'isica de Materiales, Universidad Aut\'onoma de Madrid, UAM, 28049 Madrid, Spain}
\altaffiliation{Instituto de Ciencia de Materiales ``Nicol\'as Cabrera'', Campus de Cantoblanco, E-28049 Madrid, Spain }
\author{Isabel J. Ferrer}
\affiliation{Materials of Interest in Renewable Energies Group (MIRE Group), Dpto. de F\'isica de Materiales, Universidad Aut\'onoma de Madrid, UAM, 28049 Madrid, Spain}
\altaffiliation{Instituto de Ciencia de Materiales ``Nicol\'as Cabrera'', Campus de Cantoblanco, E-28049 Madrid, Spain }

\author{Gabino Rubio-Bollinger}
\affiliation{Dpto. de F\'isica de la Materia Condensada, and Condensed Matter Physics Center (IFIMAC), Universidad Aut\'onoma de Madrid, Campus de Cantoblanco, E-28049 Madrid, Spain}
\altaffiliation{Instituto de Ciencia de Materiales ``Nicol\'as Cabrera'', Campus de Cantoblanco, E-28049 Madrid, Spain }

\author{Andres Castellanos-Gomez}
\email{andres.castellanos@imdea.org}
\affiliation{Instituto Madrile\~no de Estudios Avanzados en Nanociencia (IMDEA Nanociencia), Campus de Cantoblanco, E-28049 Madrid, Spain.}

\author{Roberto D'Agosta} 
\email{roberto.dagosta@ehu.es}
\affiliation{Nano-Bio Spectroscopy Group and European Theoretical Spectroscopy Facility (ETSF), Universidad del Pa\'is Vasco, E-20018 San Sebasti\'an, Spain}
\altaffiliation{IKERBASQUE, Basque Foundation for Science, E-48013, Bilbao, Spain}
 
\date{\today} 
\begin{document}

\begin{abstract} 
By combining {\it ab initio} calculations and experiments we demonstrate how
the band gap of the transition metal tri-chalcogenide TiS$_3$ can be modified
by inducing tensile or compressive strain. We show by numerical calculations
that the electronic band gap of layered TiS$_3$ can be modified for monolayer,
bilayer and bulk material by inducing either hydrostatic pressure or strain. In
addition, we find that the monolayer and bilayer exhibits a transition from a
direct to indirect gap when the strain is increased in the direction of easy
transport. The ability to control the band gap and its nature can have an
impact in the use of TiS$_3$ for optical applications. We verify our prediction
via optical absorption experiments that present a band gap increase of up to 10\%
upon tensile stress application along the easy transport direction.
\end{abstract}

\maketitle 

%\begin{document} 
The discovery of graphene \cite{Novoselov2004} paved the way
to the investigation to a vast class of materials whose salient characteristic
is to exist in single or a few layers. Similar to graphene for example, the
existence of silicene and germanene has been predicted \cite{Cahangirov2009,
Yang2014,Cahangirov2013}. At the same time, the mechanical exfoliation
technique that allowed the breakthrough for the carbon based material, going
from graphite to graphene, allows to produce single (or a few) layers of other
materials. A family of these materials are the transition metal dichalcogenides
(TDMs) like, e.g., MoS$_2$, MoSe$_2$, WS$_2$, and WSe$_2$. TDMs have
shown semiconducting properties, like a direct gap and tunability, superior to
certain extent to those of graphene and could form a basis for novel
transistors and photodetectors. Ideally, these materials should have a gap
comparable with that of Silicon to ease the integration on the existing
technology. However, TDMs show a direct gap only when reduced to a single
layer, due to interlayer interaction. This might limit their systematic use. On
the other hand, trichalcogenides can be reduced to single or a few layer
devices while showing a direct gap of about 1 eV \cite{Endo1981}. In particular
titanium trisulphide, TiS$_3$, shows an extremely fast optical response: an
ideal property for the next generation of photodetectors
\cite{Island2014,Island2015,Molina-Mendoza2015}. Moreover, it has been predicted
that TiS$_3$ could be a promising electrode material for Li and Na in batteries
\cite{Wu2015}, as photoelectrode for H2 photogeneration \cite{Barawi2015}, and nano-electronics and optics \cite{Jin2015}, presenting also
strong anisotropic behavior and non-linearity both in the electronic and optical properties
\cite{Gorlova2012,Island2015,Molina-Mendoza2015,Dai2015}. 

The exfoliation of a few layer TiS$_3$ is a relatively recent achievement
\cite{Island2014}. It has also been shown that by controlling the growing
conditions one can obtain different materials, from nano-ribbons to flakes as
well as control several electrical properties of TiS$_3$. In particular, the
presence of S vacancies has an important effect on the electronic transport
properties of the few layers nano-ribbons \cite{Island2015,Iyikanat2015}. In
this Letter we show that one can control the electronic band gap by inducing compressive and expansive uniaxial strain or hydrostatic pressure to the material.
Here, we consider the cases of bulk, mono- and bi-layer devices, and show that
one can induce a direct-to-indirect band-gap transition by inducing strain into
the electronic transport facile axis. We verify this prediction by stretching a
thin TiS$_3$ sample and by measuring the band-gap via optical absorption spectroscopy.

To calculate the electronic band structure, we have performed state-of-the-art
DFT calculations with a pseudo-potential plane-wave
method as implemented in the PWSCF code of the Quantum-ESPRESSO suite
\cite{Giannozzi2009}. For both Ti and S, the electron exchange-correlation
potential is evaluated within the generalized gradient approximation throughout
the Perdew-Burke-Ernzerhof's functional. For S the Martins-Troulliers'
pseudo-potential is used, while for Ti the Goedecker-Hartwigsen-Hutter-Teter's
pseudo-potential, including semi-core states for the valence electrons, is used
\cite{Hartwigsen1998,Goedecker1996}. These pseudo-potentials are
norm-conserving and scalar relativistic. By starting from the experimental
parameters for the unit cell \cite{Island2015} and the spectroscopical atomic
configuration \cite{Furuseth1975}, we have optimized the atomic positions with
a residual force after relaxation of $0.001$ a.u. using the
Broyden-Fletcher-GoldfarbShann's procedure. 
The kinetic energy cutoff for the plane wave basis set is put at $220$ Ry,
while the cutoff for the charge density is $880$ Ry. The sampling of the
Brillouin zone for the bulk material is $10\times 10 \times 10$ according to
the Monkhorst-Pack scheme, while for the mono- and bi-layer material we used a
$\bf{k}$-point mesh of $14\times 14 \times 4$. The parameters chosen ensure a
convergence of the band gap within an accuracy of around $0.01$ eV. We have not
included van der Waals corrections in these calculations since from our
previous experience we have seen they have essentially no effect on the
electronic bands.
\begin{figure}[ht!]
\includegraphics[width=8cm]{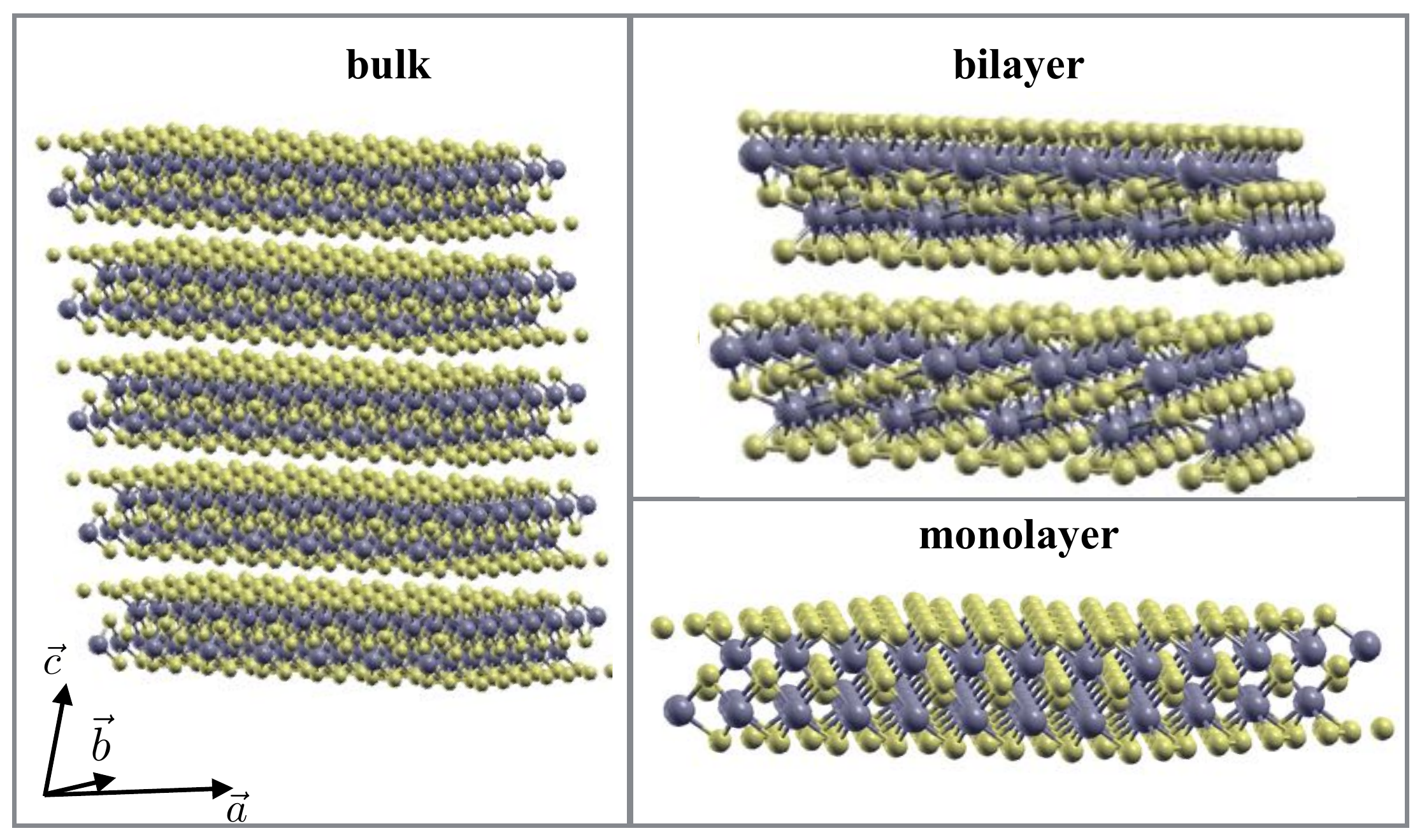} 
\caption{
Sketch of the three different atomic structures under investigation. Bulk
TiS$_3$ is a layered material where the different layers are essentially weakly
interacting. The unit cell vector for the bulk are shown in the left figure.
For monolayer and bilayer structures the $\vec a$ and $\vec b$ vectors are the
same.}
\label{structure} 
\end{figure} 

We have calculated the electronic band structure of mono- and bi-layer, and
bulk TiS$_3$. We show in Fig.~\ref{structure} the atomic configurations of
these materials. For the mono- and bi-layer material we found that they are
direct gap semiconductors with a DFT gap of around 0.3 eV. On the other hand,
the bulk material is predicted as an indirect gap semiconductor with a DFT gap
of around 0.3 eV. In contrast, the experiments indicate that bulk TiS$_3$ is a
direct gap semiconductor with an electronic gap of about 1 eV
\cite{Endo1981,Island2015}. For all the cases we have considered here, a more
refined calculation based on non-self-consistent GW method (or hybrid
functionals) opens up the gap to the experimental levels (about 1.0 eV),
\cite{Molina-Mendoza2015} but does not remove the discrepancy on the nature of
the gap (these results are not shown here). From our work \cite{Island2015}
with different pseudo-potentials it seems that a pivotal role in determining
the nature of the gap in bulk TiS$_3$ is played by the core electrons of the Ti
atoms and in particular how they are described by the exchange-correlation
potential used in the DFT calculation. We believe that this discrepancy between
theory and experiment is still open and further investigation is necessary. On
the other hand, we expect the \emph{nature} of the gap does not play a crucial role on the
way we control it via strain or pressure.

\begin{figure}[ht!]
	\includegraphics[width=8.0cm]{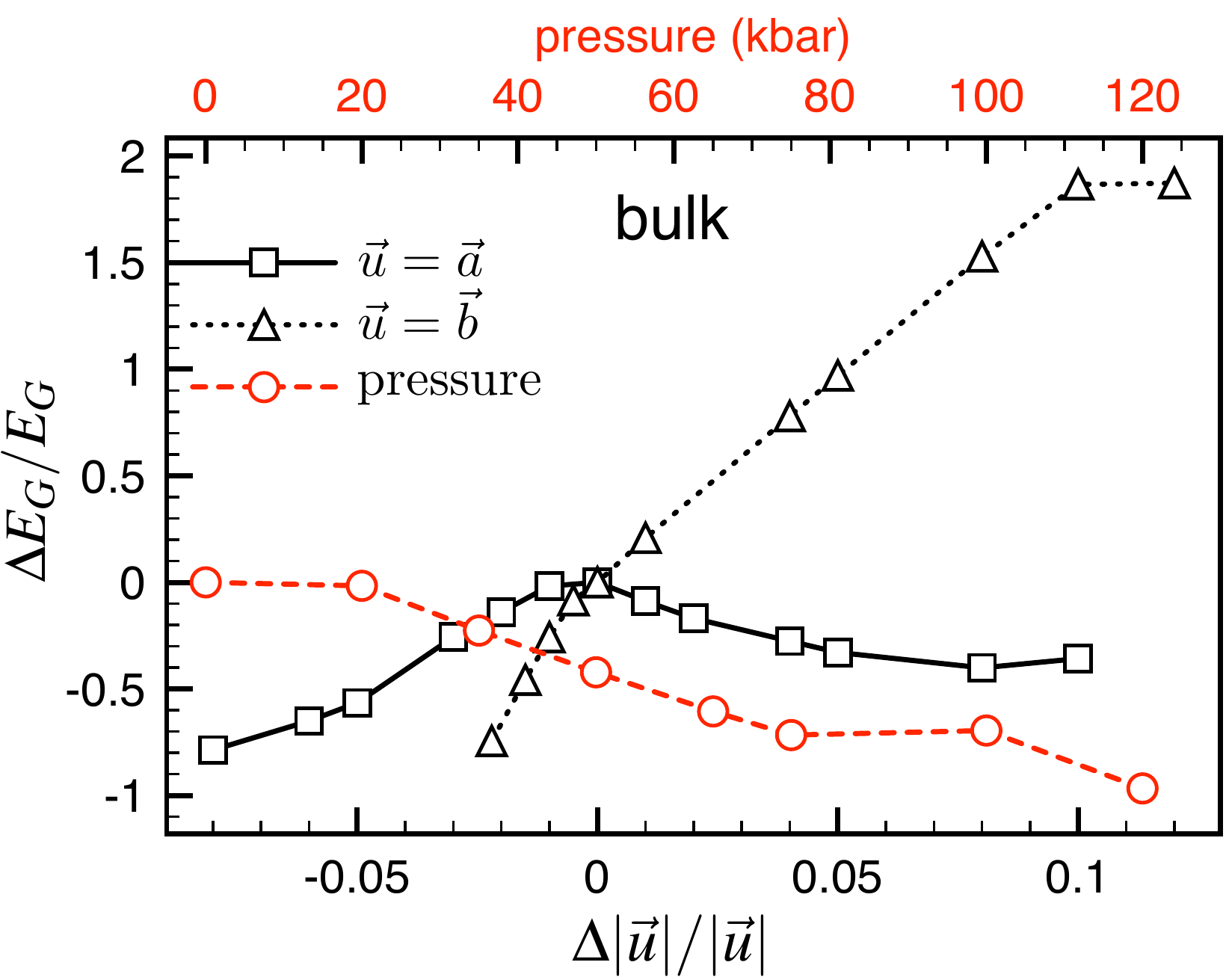}
	\caption{ Modulation of the indirect band gap, $\Delta E_G$ of bulk TiS$_3$ by inducing strain or applying pressure. Pressure reduces the gap until it closes at around $120$ kbar (red dashed line). Any strain in the $\vec a$ direction reduces the gap (black solid line), while strain in the $\vec b$ direction can leads to an increase or reduction of the gap (doted black line). Here, $E_G$ is the band gap of bulk TiS$_3$ when strain or pressure are not applied.}
	\label{bulk_strain}
\end{figure}

In order to study the strain induced band gap modulation, we applied a stress
by deforming the length of the unit-cell vector in the $\vec a$ or $\vec b$
direction and relaxed the atomic positions.\footnote{We here neglect the effect
of the Poisson's deformation induced in the other two axes of the unit cell.
Preliminary results have shown this effect to be negligible with Poisson's
ratios of the order of 0.04-0.08. These results will be presented elsewhere.}
In Fig. \ref{bulk_strain} one can see the indirect band gap change $\Delta E_G$
for bulk TiS$_3$. $E_G$ refers to the band gap of the material when pressure or
strain are not applied. While a negative strain (compression) in $\vec a$ and
$\vec b$ direction of the unit cell leads to a reduction of the band gap, a
positive strain can increase or decrease the band gap. A positive strain in the
$\vec a$ direction leads to an increase of the indirect band gap, while in the
$\vec b$ direction the positive strains induce a reduction of the electronic
gap. In addition, we have studied the influence of an isotropic stress,
hydrostatic pressure, on the bulk material. Pressure leads to a reduction of the band gap,
which coincides with the results that negative stress in both $\vec a$ and
$\vec b$ directions lead to a reduction of the gap. These findings are quite
unique since for TDMs a tensile strain always reduces the band gap
\cite{Roldan2015}.

\begin{figure}[ht!]
	\includegraphics[width=8.0cm]{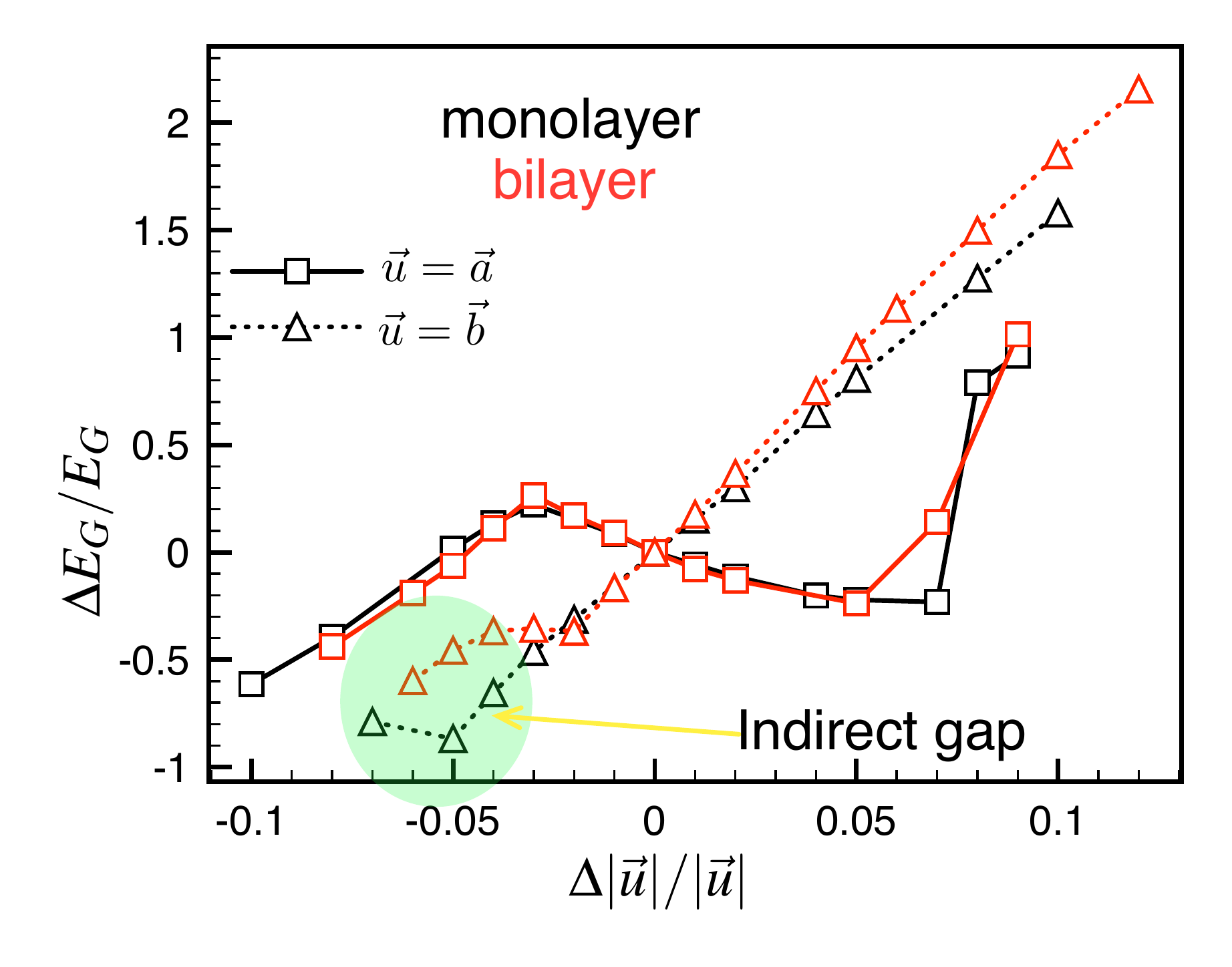}%{_mono_strain}
	\caption{ Modulation of the direct band gap of mono- (black) and bi-layer (red) TiS$_3$ by applying tensile or compressive strain, along the unit cell vector $\vec a$ or $\vec b$. All gaps are direct except the green shaded area: A direct to indirect gap transition is predicted when negative strain is applied to the $\vec b$ direction at around $-4 \% $.}
	\label{mono_strain}
\end{figure}

While for bulk we only find a reduction or increase of the gap, for monolayer
and bilayer TiS$_3$ we find additionally a direct-indirect transition, this
might have important consequences to tailor the optical response of possible
TiS$_3$ based phototransistor. In Fig. \ref{mono_strain} the influence of strain on the DFT gap of
mono- and bi-layer TiS$_3$ is shown. For both systems, while a compression in the $\vec a$ direction
reduces the direct gap, an expansion increases the gap until a strain of
around $-3 \%$ and increasing the negative strain further leads to a
reduction of the gap. Most interesting, a compression in $\vec b$ direction
leads to a reduction of the direct gap up to $-3 \% $, applying $-4 \% $ or
more strain transforms the gap to an indirect one by keeping the gap
constant at around one third of its original value. As in the bulk material a
positive strain in the $\vec b$ direction leads to an monotonic increase of the
gap. 

\begin{figure}[t!]
	\includegraphics[width=8.0cm]{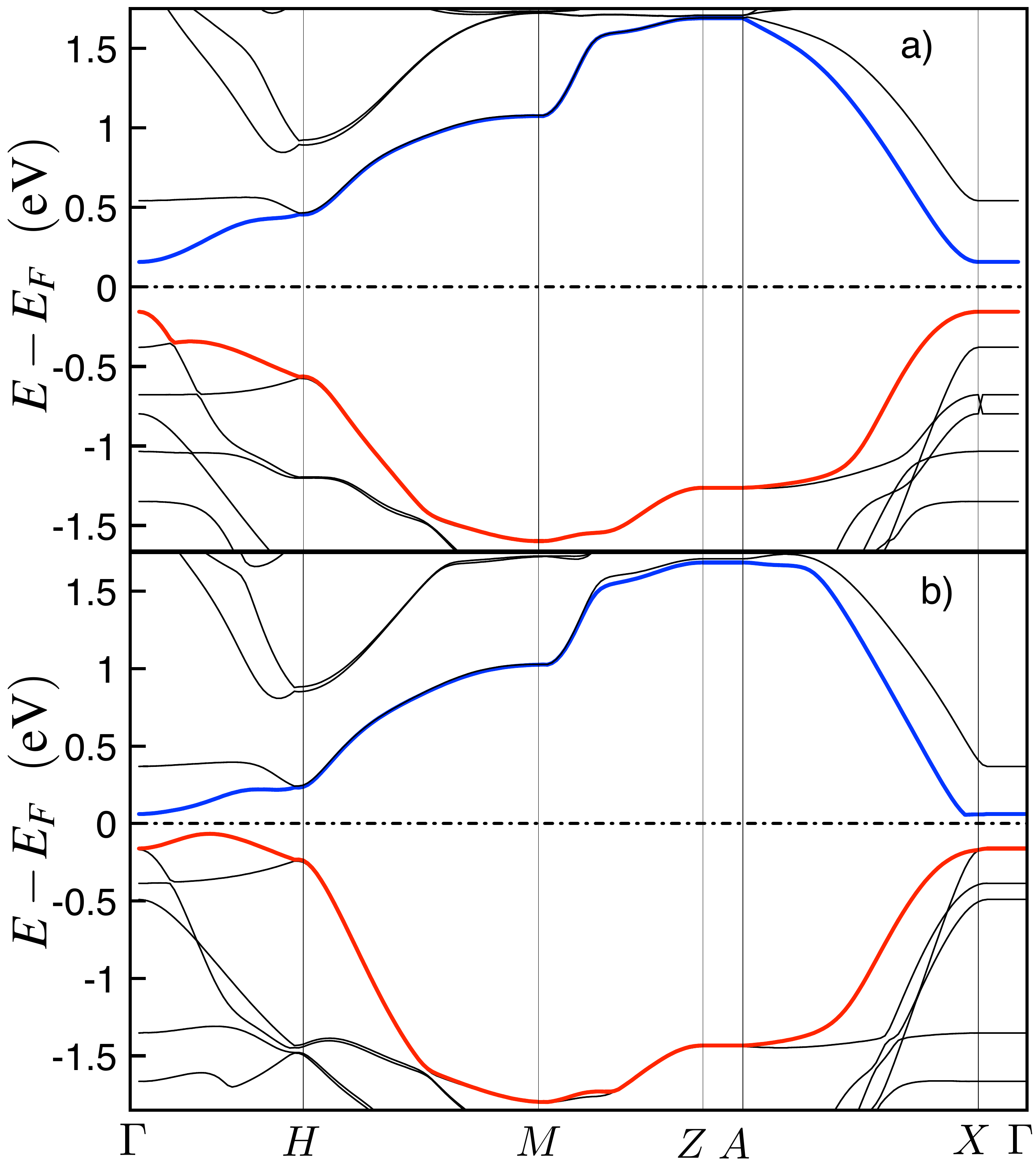}\\%bands_mono_0}\\
	\caption{ Electronic bands for the monolayer TiS$_3$ for the relaxed cell structure
(top) and the one where we compressed $5\%$ the unit cell in the $\vec b$
direction (bottom). In blue and red we represent the conduction and valence band, respectively. a) The upper plot, corresponding to the relaxed unit cell,
shows an direct gap at the $\Gamma$-point. b) The compressed structure shows
a decreased indirect gap between $\Gamma$ and $H$ The compression strongly modifies the valence band between these two points. All the energies have been rescaled to the Fermi energy $E_F$.}
	\label{mono_bands}
\end{figure}

To better understand this transition, the band structure for the unstrained
monolayer and 5\% strain in the $\vec b$ direction is shown in Fig.
\ref{mono_bands}. We clearly see from Fig. \ref{mono_bands} that the conduction
bands is almost untouched by the compression, while the valence bands develops
a local maximum between $\Gamma$ and $H$, reducing the gap and changing its
nature. Interestingly the reduction of the band-gap in the transition from
direct to indirect is here more marked than what has been reported in TiS$_3$
nano-ribbons, where the reported band gap modification is of only a few meV
\cite{Kang2015}, in certain cases below the accuracy of the pseudo-potentials
used in DFT.

\begin{figure}[ht!]
\includegraphics[width=8cm]{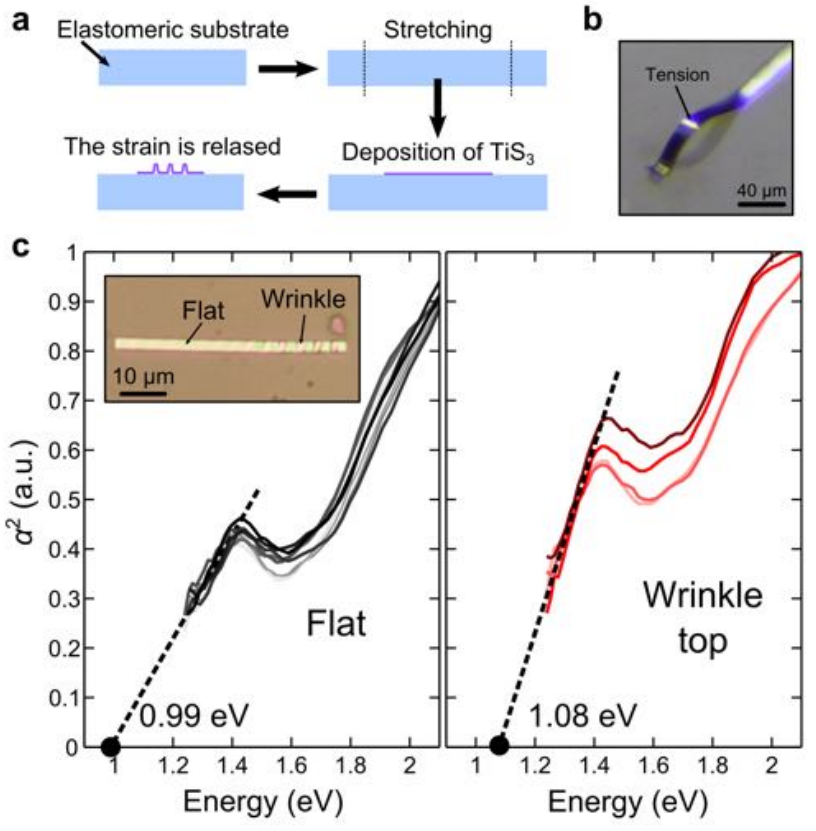}
\caption{ (a) Schematic diagram of the steps employed to
fabricate uni-axially strained TiS$_3$ exploiting the buckling-induced
de-lamination process. (b) High-angle optical microscopy image of a delaminated
wrinkle occurring on a thin TiS$_3$ ribbon. (c) Optical absorption spectra
acquired on 9 different flat regions (left) and on the topmost part of 4
wrinkles on a thin TiS$_3$ ribbon (shown in the inset). The intercept with the
horizontal axis indicates the estimated band-gap value in the different
regions.}
    \label{experiments}
\end{figure}

In order to experimentally verify the predicted effect of strain on the
electronic band structure of TiS$_3$ we study the optical absorption spectra of
a thin TiS$_3$ ribbon subjected to uniaxial strain. The strain is applied by
exploiting the buckling-induced de-lamination process that takes place when a
thin elastic film, deposited onto an elastomeric substrate, is subjected to a
uniaxial compressive strain \cite{Castellanos-Gomez2013,Yang2015}. The
trade-off between the bending rigidity of the thin-film and the
thin-film/elastomeric substrate adhesion forms wrinkles that delaminate from
the elastomeric substrate where the thin-film is uni-axially stretched. We
address the reader to Ref. \cite{Mei2011} for more details on the
buckling-induced de-lamination process. Figure \ref{experiments}a shows a
sketch of the process followed to fabricate the uni-axially strained TiS$_3$
sample. A gelfilm substrate (a commercially available elastomeric substrate) is
uni-axially stretched by 30$\%$, then TiS$_3$ is deposited onto the stretched
surface and the strain is suddenly released yielding to the buckling-induced
de-lamination to the TiS$_3$ (see Figure \ref{experiments}b) with flat regions
(released stress) and delaminated wrinkles (accumulated tensile stress). For
thin TiS$_3$ ribbons (10 nm to 30 nm thick, like the one studied here) the
wrinkles are 100 nm - 300 nm in height, the estimated maximum tensile strain on
the topmost part is in the order of 0.3-0.7 $\%$
\cite{Castellanos-Gomez2013,Vella2009}. The change in the band structure
induced by the applied uniaxial strain along $\vec b$ is probed by a recently
developed hyper-spectral imaging based absorption spectroscopy technique. We
address the reader to Ref. \cite{Castellanos-Gomez2015} for details on this
technique. Figure \ref{experiments}c shows the absorption spectra $\alpha^2$
acquired on 9 flat regions and on top of 4 wrinkles. The intercept with the
horizontal axis of the relationship $\alpha^2$ vs. $E$ gives an estimate of the
band gap (valid for direct gap semiconductors, for indirect ones it should be
$\alpha^{1/2}$ vs. $E$). The obtained band gap values in the flat region, 0.99
eV, is in good agreement with the value determined by conventional absorption
spectroscopy and by photocatalysis measurements on bulk material
\cite{Molina-Mendoza2015,Ferrer2013}. On the topmost part of the wrinkles the
slope of the absorption band edge increases considerably yielding an estimated
band gap value of 1.08 eV, 90 meV higher than on the unstrained TiS$_3$. This
experimental observation proves that 0.3-0.7 \% uniaxial tensile strain along
$\vec b$ locally changes the band structure opening the band gap, without
changing the nature of the band-gap which remains direct (as evidenced by the
marked linear behavior of the absorption band edge in the representation
$\alpha^2$ vs. $E$). The experimental value also confirms the quite high
sensitivity of these samples as predicted by our calculations: From Fig.
\ref{bulk_strain} one can infer for small amplitude strains the linear relation
$\Delta E_G/E_G\simeq 20~\Delta |\vec u|/|\vec u|$ for strains in the $\vec b$
direction in relatively good agreement with the experimental value 13-30. Note that the large uncertainty of the experimentally determined strain tunability stems from the difficulty in estimating the actual strain applied to the sample.

In conclusion, we have done a systematic investigation of the band gap
modifications of monolayer, bilayer and bulk TiS$_3$ material when the system
is subject to strain or hydrostatic pressure. We found at zero applied strain or pressure an indirect
gap for the bulk of around 0.28 eV, while for the mono- and bi-layer material
we calculated a direct of 0.31 eV and 0.27 eV, respectively. These values,
although in good agreement with other theoretical calculations, are about 1/3 of the experimental band gap value, a not surprising results considered that DFT usually underestimate the electronic band gap.
 We found that the gap can be controlled by inducing strain in
certain directions along the primitive axes of the unit cell. Most
interestingly, we found that the gap of the monolayer and bilayer material
changes from direct to indirect when compressive strain is induced along the preferred
transport axis ($\vec b$ in Fig. \ref{structure}). We tested our predictions by
inducing strain in a TiS$_3$ sample: the gap increases when the system is
under tensile strain, in agreement with the theoretical predictions of
Fig.~\ref{bulk_strain}. This behavior is mostly noticeable, since for other two
dimensional di-calchogenides materials the gap always closes for tensile
strain, and can open a set of potential application for TiS$_3$. 
Note added: During the elaboration of the manuscript the authors became aware
of two theoretical works where, the general properties of the MX$_3$ materials
(where M = Ti, Zr and Hf; while X = S, Se, Te) \cite{Abdulsalam2015}, the
effects of tensile a strain on the band gap of single layer \cite{Li2015a} and
nano-ribbons \cite{Yang2015} of TiS$_3$ are studied.

\begin{acknowledgement}
%\acknowledgments
R.B. and R.D'A. acknowledge financial support by DYN-XC-TRANS (Grant No.
FIS2013-43130-P), the Grupo Consolidado UPV/EHU del Gobierno Vasco (IT578-13),
and NANOTherm (CSD2010-00044) of the Ministerio de Economia y Competitividad (MINECO).
R. B. acknowledges the financial support of Ministerio de Educacion, Cultura y
Deporte (FPU12/01576). R. D'A. is grateful to the Physics Department of King's
College London for its hospitality during the completion of this work supported
by the Grant No. MV-2015-1-17 of the Diputacion Foral de Guipuzkoa. A.C.-G.
acknowledges financial support from the BBVA Foundation through the fellowship
``I Convocatoria de Ayudas Fundacion BBVA a Investigadores, Innovadores y
Creadores Culturales'' (``Semiconductores ultradelgados: hacia la
optoelectronica flexible''). G. R.-B. acknowledges financial support from the Grant No. MAT2014-57915-R from the MINECO.  The MIRE group acknowledge financial support from MINECO (MAT2011-22780) and E.F. thanks to the Mexican National Council for
Science and Technology (CONACyT) for providing a PhD. Grant.
\end{acknowledgement}
\bibliography{library.bib}
\end{document}